\documentclass[aps,twocolumn,tightenlines,nofootinbib,superscriptaddress]{revtex4-2}
\usepackage[utf8]{inputenc}
\usepackage{graphicx}
\usepackage{subcaption}
\usepackage{mwe}
\usepackage{mathrsfs}
\usepackage{amsfonts}
\usepackage{setspace}
\usepackage{cellspace}
\usepackage{amsmath,amssymb,bm}
\usepackage[colorlinks=true,linkcolor=blue]{hyperref}
\usepackage{xcolor}
\usepackage{epsfig}
\usepackage{slashed}
\usepackage{caption}
\usepackage{hhline,multirow,tabularx}  
\usepackage{dcolumn}    
\usepackage{url}        
\usepackage{braket}     
\usepackage{makecell}

\renewcommand{\bra}[1]{\left<#1\left|}
\renewcommand{\ket}[1]{\right|#1\right>}

\newcommand{\ui}[1]{^{(#1)}}
\newcommand{\btheta}{{\boldsymbol{\theta}}}
\newcommand{\CalP}{{\mathcal{P}}}

\newcommand{\by}{\mathbf{y}}

\DeclareMathAlphabet{\mathdutchcal}{U}{dutchcal}{m}{n}
\SetMathAlphabet{\mathdutchcal}{bold}{U}{dutchcal}{b}{n}
\DeclareMathAlphabet{\mathdutchbcal}{U}{dutchcal}{b}{n}

\begin{document}

\title{Bayesian Inferring Nucleon's Gravitation Form Factors via Near-threshold $J/\psi$ Photoproduction}

\author{Yuxun Guo}
\email{yuxunguo@lbl.gov}
\affiliation{Physics Department, University of California, Berkeley, California 94720, USA}
\affiliation{Nuclear Science Division, Lawrence Berkeley National
Laboratory, Berkeley, CA 94720, USA}

\author{Feng Yuan}
\email{fyuan@lbl.gov}
\affiliation{Nuclear Science Division, Lawrence Berkeley National
Laboratory, Berkeley, CA 94720, USA}

\author{Wenbin Zhao}
\email{wenbinzhao@lbl.gov}
\affiliation{Physics Department, University of California, Berkeley, California 94720, USA}
\affiliation{Nuclear Science Division, Lawrence Berkeley National
Laboratory, Berkeley, CA 94720, USA}

\begin{abstract}
With Bayesian inference, we investigate the impact of recent near-threshold $J/\psi$ production measurements by the $J/\psi$ 007 experiment and GlueX collaboration on the extraction of proton's gravitational form factors. We apply the generalized parton distribution framework at the next-to-leading order and demonstrate a stable expansion for the near-threshold kinematics. We find that the experimental constraints are in good agreement with the state-of-the-art lattice simulations, where negative $C_q(t)$ and $C_g(t)$ are strongly preferred. This highlights a great potential to extract them from future high-precision experiments.
\end{abstract}
 
\maketitle

\textbf{\textit{  Introduction.}} 
Nucleon tomography encoded in the gravitational form factors (GFFs) has drawn rising interest in the past decade. They hold the keys to understanding the fundamental structure, including the mass, spin, and mechanical properties, of the nucleon~\cite{Ji:1994av, Ji:1996ek, Polyakov:2002yz,Polyakov:2018zvc,Burkert:2018bqq,Duran:2022xag,Burkert:2023wzr}. One of the most important questions is to identify the gluonic components, accessing which, nevertheless, has been challenging due to their elusive nature. It has been suggested long ago that heavy quarkonia can help to unveil the underlying dynamics of the gluon~\cite{Voloshin:1978hc,Gottfried:1977gp,Appelquist:1978rt,Bhanot:1979vb,Kharzeev:1995ij,Kharzeev:1998bz}. The near-threshold heavy quarkonium photoproduction,
\begin{equation}
    \gamma+p\to J/\psi+p\ ,
\end{equation}
has been extensively investigated from both experiment and theory sides to explore the gluonic GFFs in recent years~\cite{GlueX:2019mkq,Duran:2022xag,GlueX:2023pev,Gryniuk:2016mpk,Hatta:2018ina,Hatta:2019lxo,Mamo:2019mka,Gryniuk:2020mlh,Guo:2021ibg,Guo:2023pqw,Sun:2021pyw,Sun:2021gmi,Mamo:2021krl,Mamo:2022eui,Kharzeev:2021qkd,Guo:2023pqw,Guo:2023qgu,Pentchev:2024sho}. Together with the first principle lattice calculations~\cite{Shanahan:2018nnv,Shanahan:2018pib,Hackett:2023rif,Shanahan:2018pib,Wang:2024lrm}, this will provide a unique opportunity to uncover the fundamental nucleon structure. 

One especially interesting feature of the near-threshold kinematics is that the momentum transfer is large 
and the skewness parameter $\xi$ approaches 1 in the heavy-quarkonia limit. Applying the QCD factorization in the generalized parton distribution (GPD) framework and utilizing this feature, it was found that the leading contribution to the amplitude is from the lowest moments of GPDs~\cite{Guo:2021ibg,Hatta:2021can,Guo:2023qgu}. 
This leading-moment dominance allows us to constrain 
the gluon GFFs through asymptotic expansion with large $\xi$~\cite{Guo:2021ibg,Guo:2023pqw,Guo:2023qgu}. 
%
In the GPD framework, the cross-sections of near-threshold heavy quarkonium productions read~\cite{Guo:2021ibg}:
\begin{equation}
\label{eq:xsec}
\begin{split}
\frac{d \sigma}{d t}= \frac{ \alpha_{\rm EM}e_Q^2}{ 4\left(W^2-M_N^2\right)^2}\frac{ (16\pi\alpha_S)^2}{3M_V^3}|\psi_{\rm NR}(0)|^2 |G(t,\xi)|^2\ ,
\end{split}
\end{equation}
where $t$ is the momentum transfer squared, $W$ is the center of mass energy, $M_N$ and $M_V$ represent the masses of the nucleon and heavy quarkonium, respectively. $\psi_{\rm NR}(0)$ stands for the non-relativistic wave function of the heavy quarkonium at $r=0$ in the non-relativistic QCD (NRQCD)~\cite{Bodwin:1994jh}. The hadronic matrix element $G(\xi,t)$ can be written as
\begin{align}
\begin{split}
    \label{gt2ff}
    |G(t,\xi)|^2=&\left(1-\xi^2\right)(\mathcal{H}+\mathcal{E})^2-2 \mathcal{E}(\mathcal{H}+\mathcal{E}) \\ &+\left(1-\frac{t}{4M_N^2}\right)\mathcal{E}^2 \ ,
\end{split}
\end{align}
where $\mathcal{H}=\sum_{i=q,g} \mathcal{H}_{i}$ and $\mathcal{E}=\sum_{i=q,g} \mathcal{E}_{i}$ are the Compton-like form factors (CFFs) that can be factorized in terms of the quark and gluon GPDs, $H_{q/g}$ and $E_{q/g}$~\cite{Ivanov:2004vd,Chen:2019uit,Flett:2021ghh}. For example, $\mathcal{H}_{q/g}(\xi,t)$ are written as,
\begin{align}
\begin{split}
    \mathcal{H}_{q/g}(\xi,t) \equiv \int_{-1}^{1} \text{d}x~  &C_{q/g}(x,\xi,\mu_f) H_{q/g}(x,\xi,t,\mu_f)\ ,
\end{split}
\end{align}
and similarly for $\mathcal{E}_{q/g}$. The Wilson coefficients $C_{q/g}$ have been computed up to NLO~\cite{Ivanov:2004vd,Chen:2019uit,Flett:2021ghh}. 
At LO, only gluons contribute, and the leading-moment approximation leads to $\mathcal{H}(\xi,t)\approx  2 \left[A_g(t)+4\xi^2 C_g(t)\right]/\xi^2$ and $\mathcal{E}(\xi,t)\approx  2 \left[B_g(t)-4\xi^2 C_g(t)\right]/\xi^2$ with $A_g(t)$, $B_g(t)$ and $C_g(t)$ the gluon GFFs~\cite{Ji:1996ek}. This provides an important constraint on 
the GFFs. 
To establish a solid case to extract the GFFs from these experiments, one has to go beyond the above LO picture. 
In this work, we will advance the theory developments in two aspects: apply the NLO calculations~\cite{Ivanov:2004vd,Chen:2019uit,Flett:2021ghh} in the threshold kinematics with large-$\xi$ expansion and utilize the Bayesian inference method in the phenomenological analysis of the experimental data. 

Taking into account the NLO corrections will not only help to reduce the theoretical uncertainties but also lead to access to the quark GFFs because the quark GPDs will start to contribute at the NLO. Therefore, at this order, we will be able to constrain both quark and gluon GFFs, and their total contribution to the proton's GFFs. In addition, 
to improve the accuracy of our analysis, we will employ the advanced Bayesian inference method. This offers a statistically
rigorous and interpretable extraction, which will be extremely valuable since both quark and gluon GFFs enter at NLO  and are strongly correlated in the analysis.
Advanced computational methods, such as Bayesian inference, machine learning/artificial intelligence, have been widely applied in nuclear physics, see, e.g., Refs.~\cite{Boehnlein:2021eym,Boehnlein:2025slv}, and in particular to constrain the nucleon properties related to the GPDs~\cite{Dutrieux:2024bgc,Almaeen:2024guo,Hossen:2024qwo,Adams:2024pxw}. 
In our analysis, we will also consider the constraints from the lattice simulations~\cite{Hackett:2023rif}. These lattice results will constrain, in particular, the GFFs at small momentum transfer. The combined fit to the lattice results and experimental data give us the most accurate picture of the gluon GFFs.

We also notice that the underlying mechanism for near-threshold $J/\psi$ photoproduction has been explored in different frameworks~\cite{Du:2020bqj,JointPhysicsAnalysisCenter:2023qgg,Tang:2024pky,Duan:2024hby,Wu:2024xwy}. Future experiments at JLab and the electron-ion collider will help to identify the production mechanism. 



\textbf{\textit{GPD framework at NLO in QCD.}} 
In the GPD framework, the differential cross-section of near-threshold heavy quarkonium photoproduction depends on the CFFs. As discussed in~\cite{Guo:2023qgu}, the conformal expansion of CFFs provides better convergence in the large-$\xi$ kinematics, which we will focus on in the following analysis. 
The gluonic CFFs (gCFFs) can be formally written in terms of the conformal moments of GPD as,
\begin{align}\label{eq:gcffreconf}
\begin{split}
    \text{Re}\mathcal{H}_{g}(\xi,t)= \sum_{n=0}^{\infty}\frac{1}{\xi^{2n+2}} \bar{C}\ui{2n+1}_g \bar{H}_g\ui{2n+1}(\xi,t)\ ,
\end{split}
\end{align}
which also applies to the quark CFF (qCFF) and the $\mathcal{E}_{q,g}$ CFFs. The basis $p_{q/g}^j(x,\xi)$ and its dual $c_{q/g}^j(x,\xi)$ of this conformal expansion with conformal spin $j$ are orthonormal: $\int_{-1}^{1} \text{d}x~ c_{i}^j(x,\xi)p_{i}^k(x,\xi) =(-1)^j \delta^{jk}$ for $i=q,g$~\cite{Mueller:2005ed}, such that the conformal moments of GPDs are,
\begin{equation}
    \bar{H}_{q/g}\ui{2n+1}(\xi,t) =\int_{-1}^{1}\text{d}x~H_{q/g}(x,\xi,t) c_{q/g}^{2n+1}(x,\xi)\ ,
\end{equation}
and the corresponding Wilson coefficients are
\begin{equation}
    \bar{C}_{q/g}\ui{2n+1} =\pm \int_{-1}^{1}\text{d}x~C_{q/g}(x,1) p_{q/g}^{2n+1}(x,1)\ .
\end{equation}
As the Wilson coefficient $C_{q/g}(x,\xi)$ depends solely on $x/\xi$ up to an overall factor, its $\xi$-dependence can be pulled out and $\xi$ is set to 1 in the above expression. A conventional $-$ sign appears in the prefactor for the quark. For heavy vector meson, only the singlet quark GPD $H_{\Sigma}(x,\xi,t)\equiv\sum_{\substack{q=u,d,\cdots}} H_q(x,\xi,t)-H_q(-x,\xi,t)$ contributes, and the corresponding Wilson coefficient $C_{\Sigma}(x,\xi)$ is normalized such that $\mathcal{H}_\Sigma=\sum_q \mathcal {H}_q$. In the following, the $q$ will always refer to the singlet $\Sigma$.

Analytical expressions of the Wilson coefficients in conformal spin space are generally not guaranteed beyond LO. For the leading moment with $j=1$, they read
\begin{align}
    \bar{C}_g\ui{1}&= \frac{5}{4} + \alpha_S\left[\bar{c}_g^1-\frac{55}{16\pi}\log\left(\frac{m_c^2}{\mu_F^2}\right)\right]+\mathcal{O}(\alpha_S^2) \ , \label{eq:e8}\\
    \bar{C}_q \ui{1} &= 0 + \alpha_S\left[\bar{c}_q^1+\frac{10}{9\pi}\log\left(\frac{m_c^2}{\mu_F^2}\right)\right] +\mathcal{O}(\alpha_S^2) \ , \label{eq:e9}
\end{align}
calculated with the NLO Wilson coefficients~\cite{Ivanov:2004vd,Chen:2019uit,Flett:2021ghh}. The quantities in front of the $\log\left(m_c^2/\mu_F^2\right)$ are nothing but the LO anomalous dimensions~\cite{Muller:2013jur} and LO Wilson coefficients combined
. 
On the other hand, the two genuine NLO scalars $\bar{c}_g^1$ and $\bar{c}_q^1$ are complicated, of which only the numerical values are presented here:
\begin{equation}
    \bar{c}_g^1 \approx -0.369 \quad \text{and} \quad \bar{c}_q^1 \approx -0.891\ .
\end{equation}
With these Wilson coefficients, the NLO correction in the leading-moment approximation can be included in the phenomenological analysis. 

For clarity, we write the moments and Wilson coefficients as two-element vectors in the $(q, g)$ space and the evolution operator $\mathcal{E}$, particularly, will be $2\times2$ matrices. 
Then the real part of CFFs can be written as,
\begin{equation}
\begin{split}\label{eq:gcffreconfLeadMom}
    \text{Re}\mathcal{H} \approx \xi^{-2} \boldsymbol{\bar{C}}\ui{1}(\mu_F,m_c) \boldsymbol{\mathcal{E}}\ui{1}(\mu_F,m_c) \boldsymbol{\bar{H}}\ui{1}(m_c)\ ,
\end{split}
\end{equation}
with initial scale $\mu_0=m_c$ and other dependence suppressed. We also define the evolved Wilson coefficients,
\begin{equation}
   \boldsymbol{\bar{C}}^{\rm{evo}}\equiv \alpha_S(\mu_F) \boldsymbol{\bar{C}}\ui{1}(\mu_F,m_c) \boldsymbol{\mathcal{E}}\ui{1}(\mu_F,m_c)\ ,
\end{equation}
including the LO $\alpha_S$ in the cross-section formula in eq. (\ref{eq:xsec}) which also depends on $\mu_R$(=$\mu_F$). Setting $\mu_F=m_c$ eliminates the evolution and resums the $\log(m_c^2/\mu_F^2)$ in the Wilson coefficients. Here, we can instead estimate the theoretical uncertainties beyond NLO by varying $\mu_F$. 

\begin{figure}[h]
    \includegraphics[width=0.4\textwidth]{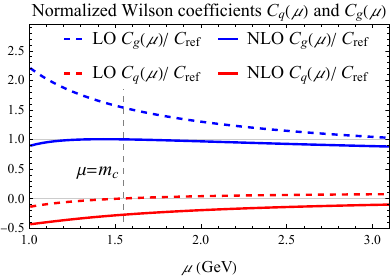}
    \caption
    {\raggedright The $\mu_F$-dependence of the LO and NLO quark and gluon Wilson coefficients $\bar{C}^{\rm{evo}}_{q}(\mu_F)$ and $\bar{C}^{\rm{evo}}_{g}(\mu_F)$, normalized by $C_{\rm{ref}}=\bar{C}_{g}(\mu_c)$ at NLO. The dependence on $\mu_F$ gets much weaker when including NLO effects, implying a good perturbative convergence.}
    \label{fig:wilcoeplt}
\end{figure}

The $\mu_F$-dependence of the LO and NLO quark and gluon Wilson coefficients $\bar{C}^{\rm{evo}}_{q}(\mu_F)$ and $\bar{C}^{\rm{evo}}_{g}(\mu_F)$, normalized by $C_{\rm{ref}}=\bar{C}_{g}(\mu_c)$ at NLO is shown in Fig. \ref{fig:wilcoeplt}, utilizing the above Wilson coefficients and NLO GPD evolution code in~\cite{Guo:2023ahv,Guo:2024wxy}. 
While the scale-dependence appears relatively strong at LO, it gets significantly reduced when including the NLO effects, suggesting a good perturbative convergence beyond NLO. 
Meanwhile, Fig. \ref{fig:wilcoeplt} also helps us to estimate the theoretical uncertainty due to perturbative corrections by varying the scale $\mu_F$. 
In particular, we find that the uncertainty is reduced from $50\%$ at LO to $10\%$ at NLO for the gluon channel contribution with a scale variation from $\mu=1$ GeV to $\mu=2m_c$.

We would like to comment on two additional corrections that include
the relativistic corrections from NRQCD and 
the large-$\xi$ expansion in the near-threshold kinematics. 
As for the relativistic corrections, they are typically suppressed by $\mathcal{O}(v^2)\sim\mathcal{O}(\alpha_S^2)$ which are of order 10\% around the scale of the charm mass~\cite{Hoodbhoy:1996zg}. This relativistic correction can be calculated in the NRQCD framework~\cite{Lappi:2020ufv}. 
A major theoretical uncertainty comes from 
large-$\xi$ expansion of the real part of q/gCFFs. Previously, we studied the asymptotic behavior with LO Wilson coefficients~\cite{Guo:2023qgu}, where the truncation error can be estimated to be around 30\% for $\xi\gtrsim 0.5$ according to the superasymptotic approximation: the error of a truncated asymptotic series is given by the term of truncation. At NLO, we found that the asymptotic expansion still holds, while the error gets generally enhanced by the more singular NLO Wilson coefficients. However, since the NLO contributions are additionally suppressed by $\alpha_S$, the overall truncation error can still be estimated to be around 30\% for $\xi\gtrsim 0.5$. This theoretical uncertainty gets enhanced for lower $\xi$, and thus we will exclude data with $\xi<0.5$ in this analysis. More details can be found in the supplemental materials.

\begin{figure}[h]
    \includegraphics[width=0.38\textwidth]{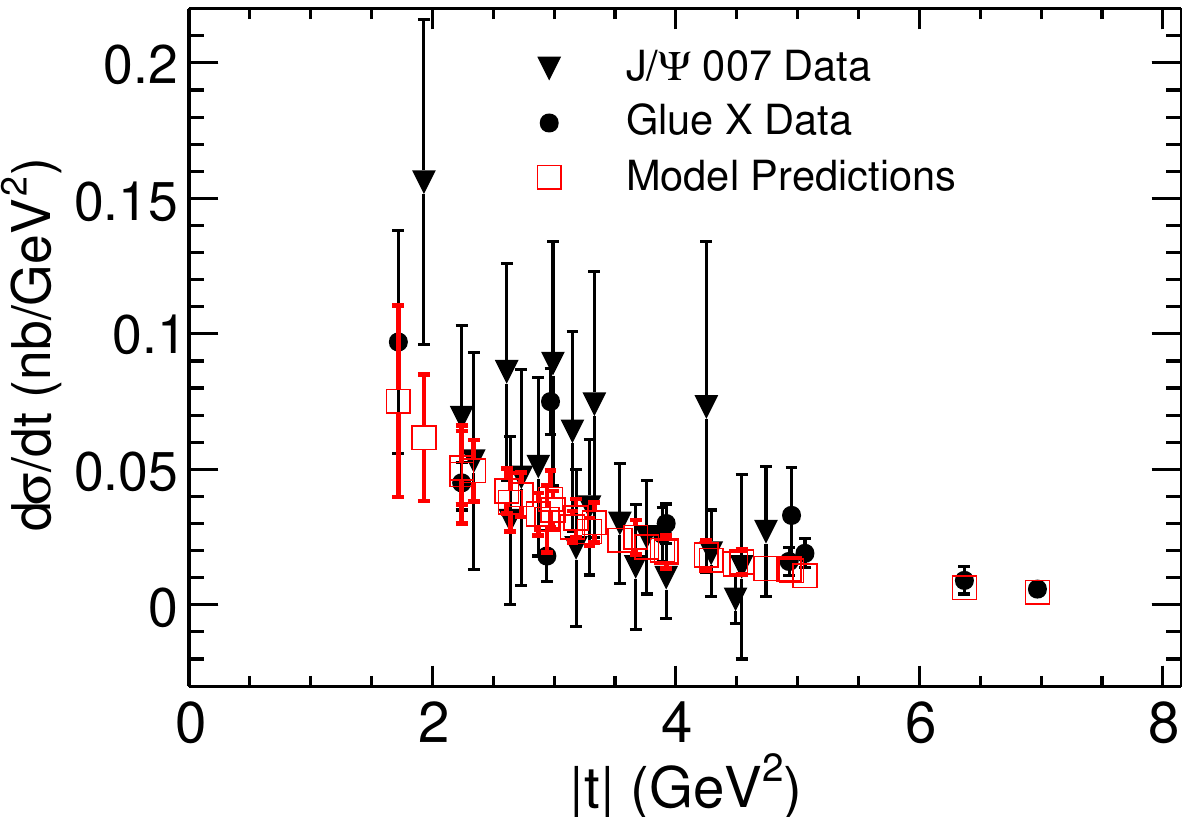}
    \caption
    {\raggedright The model predictions with the posterior distributions of the parameters fitting to experiments only with NLO accuracy. The experimental data is collected from the $J/\psi$ 007 (triangle)~\cite{Duran:2022xag} and GlueX (circle)~\cite{GlueX:2023pev} experiments across different beam energies not shown for clarity, with a cut $\xi>0.5$.}
    \label{fig:expnlo_only_comp}
\end{figure}

Accordingly, we also remark that the current limitation of constraining the GFFs from experiments is mainly due to the lack of high-precision large-$\xi$($>0.5$) data. Existing GlueX~\cite{GlueX:2019mkq,GlueX:2023pev} and $J/\psi$ 007~\cite{Duran:2022xag} data in this region contain significantly large uncertainties, see, Fig.~\ref{fig:expnlo_only_comp}. With future high-precision large-$\xi$ data, proper modeling of the higher moments contributions to the real parts of q/gCFFs as well as their imaginary parts will be necessary to improve the theoretical uncertainties.

\textbf{\textit{Phenomenological implications with Bayesian inference.}} Bayesian Inference is a general and systematic method to constrain the probability distribution of model parameters $\btheta$ by comparing model calculations $\by(\btheta)$ with experimental and lattice data $\by_\mathrm{data}$. It provides the posterior distribution $\CalP(\btheta \vert \by_\mathrm{data})$ of model parameters according to Bayes' theorem: $\CalP(\btheta \vert \by_\mathrm{data}) \propto \CalP(\by_\mathrm{data} \vert \btheta) \CalP(\btheta)$ with $\CalP(\by_\mathrm{data} \vert \btheta)$ the likelihood for model results with parameter $\btheta$ to agree with the experimental data. We choose a multivariate normal distribution for the logarithm of the likelihood~\cite{williams2006gaussian}:
\begin{equation}
    \ln[\CalP(\by_\mathrm{data} \vert \btheta)] = - \frac{1}{2} \Delta \by(\btheta)^T \Sigma^{-1} \Delta \by(\btheta)  -\frac{1}{2} \ln[(2\pi)^n \det \Sigma] ,
\end{equation}
where $\Delta \by(\btheta) = \by(\btheta) - \by_\mathrm{data}$, $n$ is the number of data points and $\Sigma \equiv \Sigma_\mathrm{data} + \Sigma_\mathrm{model}$ is the $n \times n$ covariance matrix that encodes the uncertainties of data and models~\cite{Mantysaari:2022ffw,Shen:2023awv,Jahan:2024wpj}. We include the covariance matrix of the lattice output \cite{Hackett:2023rif} and assume a diagonal form for the covariance matrix of the experimental data: $\Sigma_\mathrm{exp} = \mathrm{diag}(\sigma_1^2, \cdots, \sigma_n^2)$, combing the square of the statistical and systemic errors of each element, due to the lack of knowledge of the full covariance matrix~\cite{GlueX:2019mkq,Duran:2022xag,GlueX:2023pev}.

In the analysis, we apply the leading conformal expansion of Eqs.~(\ref{eq:e8},\ref{eq:e9},\ref{eq:gcffreconfLeadMom}) and parameterize each of the GFFs $F_{i}(t)$ with two parameters $F_{i}(0)$ and $M_{F_{i}}$ at $\mu=m_c$ as,
\begin{equation}
    F_{i}(t) = F_{i}(0)\left(1-t/M_{F_{i}}^2\right)^{-p} \ ,
\end{equation}
with $F_{i}= \{A_q,A_{g},C_{q},C_{g}\}$. Here, we neglect the contributions from  $B_{g}(t)$ and $B_{u+d+s}(t)$ form factors, which have been shown to be very small in various lattice simulations~\cite{Pefkou:2021fni,Hackett:2023rif}. We choose $p=2$ (dipole) for the $A_{q,g}$ form factors and $p=3$ (tripole) for the $C_{q,g}$ form factors based on their large-$t$ asymptotic behaviors~\cite{Sun:2021pyw,Sun:2021gmi}. The cross-sections are given by Eqs. (\ref{eq:xsec}) and (\ref{gt2ff}) with a $40\%$ theoretical uncertainty assigned to each q/gCFF as estimated above. The NRQCD matrix element is taken as $| \psi_{\rm{NR}}(0)|^2= 1.0952 /(4\pi) \left(\text{GeV}\right)^3$ for the $J/\psi$ determined from its leptonic decay width~\cite{Eichten:1995ch,Eichten:2019hbb}.

We employ the Monte-Carlo Markov Chain (MCMC) method to efficiently explore the model parameter space~\cite{goodman2010ensemble, foreman2013emcee}, and perform the analysis in three setups: one with lattice GFFs only as the baseline for comparison, one with the threshold $J/\psi$ photo-production cross-sections only, and another one combining the lattice GFFs and experiment data. The lattice data including its covariance matrix and the experimental data included in the Bayesian analysis are from \cite{GlueX:2019mkq,Duran:2022xag,GlueX:2023pev,Hackett:2023rif}. Below, the key findings are summarized, with details in the supplemental materials.



\begin{figure}[ht]
\includegraphics[width=0.48\textwidth]{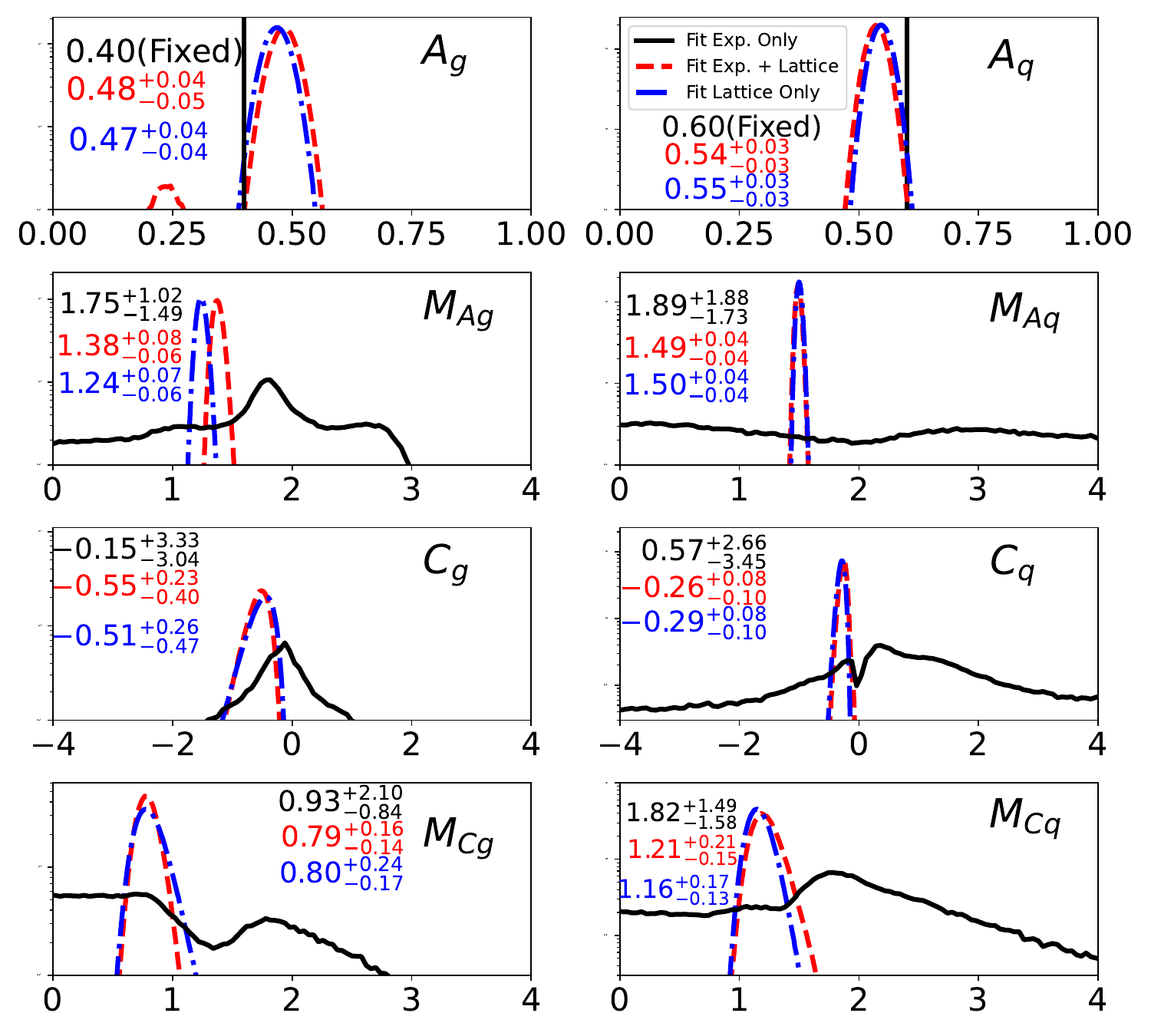}
    \caption
    {\raggedright Bayesian posterior distributions of the gluon (left) and quark (right) parameters by fitting to experiment data with NLO accuracy, lattice GFFs and their combination, respectively, with $\mu= m_c$ and masses in GeV. The parameters’ median and 90\% confidence intervals are shown. All $y$-axes are in logarithmic scale.}
    \label{fig:postrior_all}
\end{figure}

By fitting to lattice GFFs and experimental data simultaneously, we find that the two observables are generally compatible. In this fit, the lattice GFFs provide more stringent constraints on GFFs due to their much higher statistics. However, the lattice QCD results do not yet take full account of all sources of systematic error, which is likely why, e.g., the lattice GFFs give $A_g(0) \approx 0.47$, whereas the physical value is around $0.40$ according to the CTEQ global PDFs~\cite{Hou:2019efy} at $\mu= m_c$. Therefore, we also consider the extraction with experimental data alone. Obviously, the quark and gluon GFFs that both enter the cross-sections cannot be well-constrained with the limited amount of large-$\xi$ threshold $J/\psi$ production data. Here, we additionally consider the constraint $A_g(0) = 0.40 \pm0.01$ at $\mu = m_c$ from CTEQ PDFs~\cite{Hou:2019efy}, and also impose the momentum sum rule that $A_g(0)+A_{q}(0) =1$. Then, we are left with 6 parameters for the $A_{q,g}(t)$ and $C_{q,g}(t)$ form factors. These parameters are fitted to the 33 data points with $\xi>0.5$, where a comparison between the experimental data and the prediction with the Bayesian posterior distributions of the model parameters fitting to experiment data only is shown in FIG.~\ref{fig:expnlo_only_comp}.

\begin{table}[ht] 
    \def\arraystretch{1.0}
    \centering
    \caption{\raggedright Summary of parameters, their prior ranges, and the highest likelihood parameters fitting to experiment data with NLO accuracy, lattice GFFs and their combination with $\mu= m_c$ and masses in GeV.}
\renewcommand{\arraystretch}{1.0} 
    \begin{tabular}
    {>{\raggedright\arraybackslash} p{0.05\textwidth} >{\raggedright\arraybackslash} p{0.05\textwidth} >{\centering\arraybackslash} p{0.11\textwidth} >{\centering\arraybackslash} p{0.11\textwidth} >{\centering\arraybackslash} p{0.11\textwidth}} \hline \hline
    Para.  & Prior  & Exp. Only & Lat. + Exp. & Lat. Only \\
    \hline
    $A_g$      & [0, 1]   & 0.40 (Fixed)  & $\hphantom{-}0.47$  & $\hphantom{-}0.48$\\ 
    $M_{Ag}$   & [0, 4]    & $\hphantom{-}1.95$ & $\hphantom{-}1.27$  & $\hphantom{-}1.23$\\ 
    $C_g$      & [-4, 4]               & $-0.12$ & $-0.51$ & $-0.46$\\ 
    $M_{Cg}$   & [0, 4]    & $\hphantom{-}2.59$ & $\hphantom{-}0.80$& $\hphantom{-}0.83$  \\  
    $A_q$      & [0, 1]  & 0.60 (Fixed)  & $\hphantom{-}0.53$   & $\hphantom{-}0.53$ \\ 
    $M_{Aq}$   & [0, 4]    & $\hphantom{-}0.1$ & $\hphantom{-}1.48$  & $\hphantom{-}1.50$\\ 
    $C_q$      & [-4, 4]              & $\hphantom{-}0.14$  & $-0.23$ & $-0.27$\\ 
    $M_{Cq}$   & [0, 4]   & $\hphantom{-}2.59$ & $\hphantom{-}1.29$   & $\hphantom{-}1.18$ \\ 
    \hline
  \end{tabular}
  \label{tab:modelparams}
\end{table}

In TABLE \ref{tab:modelparams}, we summarize the model parameters in all three fits, and present their Bayesian posterior distribution in FIG. \ref{fig:postrior_all}. Here we emphasize the findings:
\begin{itemize}
    \item Large gluonic dipole/tripole masses are eliminated by the experimental data with Bayesian posterior distributions consistent with the lattice results, as shown in TABLE \ref{tab:modelparams} and FIG. \ref{fig:postrior_all}, although with larger uncertainties due to the limited statistics.
    
    \item Though the $C_{q,g}(0)$ form factors are not constrained as well by the experiments alone, their Bayesian posterior distributions are found to peak at values statistically consistent with the lattice ones, benefitted from the fixed $A_{q,g}(0)$. 

\end{itemize}

These illustrate the great accessibility of both the quark and gluon $C_{q,g}(t)$ form factors from such processes, even though the sensitivity to the quark only appears at NLO. Recent analyses of deeply virtual Compton scattering (DVCS) measurements~\cite{Burkert:2018bqq,Burkert:2023wzr} have demonstrated their sensitivity to the quark $C_q(t)$. A more precise determination of them with future high-precision data, including those from DVCS and near-threshold heavy quarkonium photoproduction, would be especially important and interesting. They are considered as the ``last unknown global property'' of the nucleon that carries its mechanic properties~\cite{Ji:1994av, Ji:1996ek, Polyakov:2002yz,Polyakov:2018zvc,Burkert:2018bqq,Duran:2022xag,Burkert:2023wzr}, in particular, as tensor monopole moments~\cite{Ji:2021mfb,Ji:2022exr}. The physics interpretation of these form factors have attracted attentions in the last few years~\cite{Lorce:2023zzg,Lorce:2025oot,Freese:2021czn,Freese:2021qtb,Freese:2023abr}. 
An intriguing aspect that can be derived from the GFFs 
are the proton mass and scalar radii defined through the $00$ component and trace of the energy momentum tensor of the proton as~\cite{Ji:2021mtz},
\begin{align}
\left\langle r^{2}\right\rangle_m &=6 \left.\frac{d A\left(t\right)}{d t}\right|_{t=0}-6 ~\frac{C(0)}{M_N^{2}}\ ,\label{eq:e15}\\
\left\langle r^{2}\right\rangle_s &=6 \left.\frac{d A\left(t\right)}{d t}\right|_{t=0}-18 ~\frac{C(0)}{M_N^2}\ ,\label{eq:e16}
\end{align}
respectively, where 
the contributions from $\bar C_{q,g}(t)$ form factors have been canceled out between quark and gluon. 
The physics of these radii has attracted a great interest in recent years~\cite{Miller:2018ybm,Ji:2021mtz,Wang:2021dis,Burkert:2023atx,Cao:2024zlf}. In the following, we will present the constraints from our analysis.

\begin{figure}[th]
    \centering
    \includegraphics[width=0.40\textwidth]{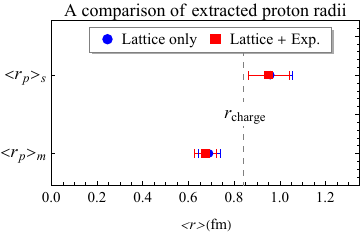}
    \caption
    {\raggedright The average proton mass and scalar radii $\left<r\right>_{s,m}\equiv \sqrt{\left<r^2\right>_{s,m}}$ with parameters sampled from their posterior distributions. Lattice GFFs taken from \cite{Hackett:2023rif}.}
    \label{fig:radius}
\end{figure}

In FIG. \ref{fig:radius}, we show the extracted proton mass and scalar radii 
from combined experimental data and lattice GFFs, compared to the ones from solely the lattice GFFs. 
Although the experimental data provides extra constraints on the proton GFFs, the extracted proton radii appear to be rather close to the lattice ones. This is partially due to the limited statistics at large-$\xi$ to be improved with future data. Finally, we report a proton scalar radius of $0.95(09)$ fm, which is comparable to the charge radius and consistent with the one from trace anomaly form factors $0.89(10)(07)$ fm~\cite{Wang:2024lrm}.
We would like to emphasize that the gluon $C_g$ terms in Eqs.~(\ref{eq:e15},\ref{eq:e16}) are the dominant source of the final uncertainties for both mass and scalar radii, and hope that future data from experiment and lattice calculation can provide further constraint.

\textbf{\textit{  Conclusion.}}
To conclude, we investigate the exclusive productions of heavy quarkonium near the threshold at NLO, which grants us the opportunity to constrain both the quark and gluon GFFs from such processes. We consider the NLO perturbative corrections of both quark and gluon Wilson coefficient as well as proper NLO GPD evolution and estimate the theoretical uncertainties.

With the above theoretical developments, we constrain the quark and gluon GFFs combining the global CTEQ PDFs~\cite{Hou:2019efy} and recent threshold $J/\psi$ production data~\cite{GlueX:2019mkq,Duran:2022xag,GlueX:2023pev} at large $\xi>0.5$ using Bayesian inference. We found that the data provide constraints on both quark and gluon $C_{q,g}(t)$ GFFs after fixing the momentum fraction $A_{q,g}(0)$ from global PDFs. We also study the corresponding proton mass and scalar radii with the extracted GFFs. This shows great potential to constrain the quark and gluon GFFs from such experiments. 
A combined analysis of both $J/\psi$ production and DVCS data will be interesting to pursue as well. We plan to carry out this in the future. 


\textbf{\textit{  Acknowledgments.}} We thank Yoshitaka Hatta for pointing out an error in $\bar C_q^{(1)}$ of Eq.(9) in an earlier version of our paper. We thank Chun Shen and Hendrik Roch for useful discussions and correspondences. We also thank Dimitra Pefkou for providing data of lattice GFFs and useful discussions and comments. We thank Xiangdong Ji and Zein-Eddine Meziani for comments and suggestions. This material is based upon work supported by the U.S. Department of Energy, Office of Science, Office of Nuclear Physics, under contract numbers DE-AC02-05CH11231, under the umbrella of the Quark-Gluon Tomography (QGT) Topical Collaboration with Award DE-SC0023646 and the Saturated Glue (SURGE) Topical
Theory Collaboration. W.B.Z. is also supported by NSF under Grant No. OAC-2004571 within the X-SCAPE Collaboration. 

\bibliographystyle{apsrev4-1}
\bibliography{refs.bib}

\clearpage
\onecolumngrid
\setcounter{page}{1} 
\setcounter{equation}{0}
\renewcommand{\theequation}{\rm{S.}\the\numexpr\value{equation}\relax}

\begin{center} \large
    \textbf{Supplemental material of ``Bayesian Inferring Nucleon's Gravitation Form Factors via Near-threshold $J/\psi$ Photoproduction''}
\end{center}

In this Supplemental Material, we provide a detailed derivation of our results presented in the Letter.

\section{Theoretical framework and setup}
As discussed in~\cite{Guo:2021ibg}, the differential cross-sections of near threshold heavy quarkonium productions reads:
\begin{equation}
\label{s.eq:xsec}
\begin{split}
\frac{d \sigma}{d t}= \frac{ \alpha_{\rm EM}e_Q^2}{ 4\left(W^2-M_N^2\right)^2}\frac{ (16\pi\alpha_S)^2}{3M_V^3}|\psi_{\rm NR}(0)|^2 |G(t,\xi)|^2\ ,
\end{split}
\end{equation}
as presented in the letter. The $G(t,\xi)$ corresponds to the hadronic matrix element that carries the properties of the nucleon. It can be written in terms of the GPDs of the nucleon, which are defined as~\cite{Ji:1998pc}
\begin{equation}
    F_{q/g}(x,\xi,t) \equiv  \int \frac{\text{d}\lambda}{2\pi} e^{i\lambda x}\bra{P'}\hat{\mathcal{O}}_{q/g}\left(\lambda n\right) \ket{P}\ ,
\end{equation}
where the momentum transfer $\Delta\equiv P'-P$ and its square $t\equiv \Delta^2$, the skewness parameter $\xi\equiv -n\cdot \Delta/[n\cdot (P+P')]$ and the average parton momentum fraction $x$ have been defined. Here $n$ is the light-like vector conjugate to the direction of $P+P'$. The non-local quark and gluon correlators are:
\begin{align}
    \hat{\mathcal{O}}_{q}\left(z\right)=\frac{1}{\bar P^+}\bar \psi\left(-\frac{z}{2}\right) \gamma^+ \psi\left(\frac{z}{2}\right) \qquad \text{and} \qquad
    \hat{\mathcal{O}}_g \left(z\right)=\frac{1}{\left(\bar P^+\right)^2}\text{Tr}\left\{ F^{+i}_{\;\;\;\;\;}\left(-\frac{z}{2}\right)F^{+}_{\;\;i}\left(\frac{z}{2}\right)\right\}\ ,
\end{align}
with implicit gauge links between the fields to maintain gauge invariance. The quark GPDs only appear at NLO and beyond when ignoring the intrinsic heavy quark. Both the quark and gluon GPDs can be further parameterized in terms of the proton spinors and scalar functions as~\cite{Ji:1998pc}:
\begin{align}
    F_{q/g}(x,\xi,t)&=\frac{1}{2\bar P^+}\bar u(P')\left[H_{q/g}(x,\xi,t) \gamma^+ +E_{q/g}(x,\xi,t)\frac{i\sigma^{+\alpha}\Delta_{\alpha}}{2 M_{N}}\right]u(P) \ ,
\end{align}
where the different Dirac structures of the nucleon are explicitly shown and $H_{q/g}(x,\xi,t)$ and $E_{q/g}(x,\xi,t)$ are the well-known leading-twist quark/gluon GPDs. Then the hadronic matrix element $G(t,\xi)$ can be written with them as,
\begin{equation}
    G(t,\xi) =\sum_{i=q,g}\int_{-1}^{1} \text{d}x~  C_{i}(x,\xi,...) F_i(x,\xi,t,\mu_F)\ , 
\end{equation}
where the Wilson coefficient $C_{i}(x,\xi,...)$ could additionally depend on the scale $m_c$, factorization scale $\mu_f$ and renormalization scale $\mu_R$ and can be calculated perturbatively. At leading-order, they read~\cite{Guo:2021ibg}
\begin{align}
    C_{g}(x,\xi)&=\frac{1}{(\xi-i0)^2-x^2} +\mathcal{O}(\alpha_S)\ , \\
    C_{q}(x,\xi)& = 0+\mathcal{O}(\alpha_S)\ .
\end{align}
The NLO results with NRQCD can be found in~\cite{Ivanov:2004vd,Chen:2019uit,Flett:2021ghh} which will be much more involving. We separately define the quark/gluon Compton-like form factors (qCFFs/gCFFs) in the hadronic matrix element as,
\begin{align}
\begin{split}
    \mathcal{H}_{q/g}(\xi,t) \equiv \int_{-1}^{1} \text{d}x~  &C_{q/g}(x,\xi,\cdots) H_{q/g}(x,\xi,t,\mu_f)\ ,
\end{split}
\end{align}
with similar definitions of $\mathcal{E}_{q/g}$ for the $E_{q/g}(x,\xi,t)$ GPDs, and the total CFFs as:
\begin{equation}
    \mathcal{H}=\sum_{i=q,g} \mathcal{H}_{i}\qquad \text{and}\qquad  \mathcal{E}=\sum_{i=q,g} \mathcal{E}_{i}\ ,
\end{equation}
so that we have eventually,
\begin{align}
    G(t,\xi)&=\frac{1}{2\bar P^+}\bar u(P')\left[\mathcal{H} \gamma^+ +\mathcal{E}\frac{i\sigma^{+\alpha}\Delta_{\alpha}}{2 M_{N}}\right]u(P) \ .
\end{align}
Summing and averaging over the polarizations of the final and initial protons, one has~\cite{Guo:2021ibg},
\begin{align}
\begin{split}
    \label{s.gt2ff}
    |G(t,\xi)|^2=\left(1-\xi^2\right)(\mathcal{H}+\mathcal{E})^2-2 \mathcal{E}(\mathcal{H}+\mathcal{E}) +\left(1-\frac{t}{4M_N^2}\right)\mathcal{E}^2 \ .
\end{split}
\end{align}
This reproduces the Eq. (3) in the main text that relate the q/gCFFs to the cross-sections.

\section{Conformal moment expansion of Compton-like amplitudes}

Another technique used in the letter is the conformal moment expansion of q/gCFFs based on the conformal moment expansion of GPDs, where detailed discussions are summarized in~\cite{Mueller:2005ed}. Here we briefly introduce this method for this material to be self-contained, taking the quark GPD $H_q(x,\xi,t)$ as an example. The construction also applies to the $E_q(x,\xi,t)$ GPD. For the gluon GPDs $H_{g}(x,\xi,t)$ and $E_{g}(x,\xi,t)$, the basis will be slightly different as presented in~\cite{Mueller:2005ed}, but the general structures remain unchanged. For a given set of orthonormal basis $p_{q/g}^j(x,\xi)$ and its dual $c_{q/g}^j(x,\xi)$, GPDs can be formally expanded as~\cite{Mueller:2005ed},
\begin{equation}
    H_q(x,\xi,t) = \sum_{j=0}^\infty (-1)^j\bar {H}_{q}\ui{j}(\xi,t)  p_{q/g}^j(x,\xi)\ ,
\end{equation}
where the conformal moments are defined by the projection of GPDs onto the $c_{q/g}^j(x,\xi)$:
\begin{equation}
    \bar {H}_{q}\ui{j}(\xi,t) \equiv \int_{-1}^1 \text{d}x H_{q}(x,\xi,t) c_{q}^j(x,\xi)\ .
\end{equation}
These conformal basis functions are essentially the analytical continuation of Gegenbauer polynomials:
\begin{equation}
   p_q^j(x,\xi)\equiv  (-1)^j\xi^{-j-1} \frac{2^j \Gamma\left(\frac{5}{2}+j\right)}{\Gamma{\left(\frac{3}{2}\right)}\Gamma(j+3)} \left[1-\left(\frac{x}{\xi}\right)^2\right] C_{j}^{\frac{3}{2}}\left(\frac{x}{\xi}\right)\quad \text{with } |x|<\xi\ ,
\end{equation}
for integer $j$, and its dual $c_q^j(x,\xi)$ can be written as,
\begin{equation}
\label{eq:conformalmoment}
  c_q^j(x,\xi)\equiv  \xi^{j} \frac{\Gamma{\left(\frac{3}{2}\right)}\Gamma(j+1)}{2^j \Gamma\left(\frac{3}{2}+j\right)}C_{j}^{\frac{3}{2}}\left(\frac{x}{\xi}\right)\ ,
\end{equation}
also for integer $j$. Generally, such expansions can be done for any given orthonormal basis, here the $C_{j}^{\frac{3}{2}}\left(\frac{x}{\xi}\right)$ are chosen as the Gegenbauer polynomials with weight $\lambda=3/2$ (for gluon it will be $\lambda=5/2$). The main reason is that they are the eigenvectors of leading order evolution of quark/gluon GPDs, so the conformal moments are multiplicatively renormalizable at leading order. Plugging in the conformal expansion of GPDs into the definition of qCFFs leads to the following definition of the Wilson coefficients in the conformal moment space:
\begin{equation}
    \bar{C}_{q}\ui{j} =(-1)^j\int_{-1}^{1}\text{d}x~C_{q}(x,1) p_{q}^{j}(x,1)\ ,
\end{equation}
where the $-$ sign remains when $j=2n+1$ and the overall $\xi$-dependence is pulled out. For gluon GPDs, there will be a  mismatch in $j$ as the gluon GPDs $H_g(x,\xi,t)$ reduces to $x g(x)$ in the forward limit with $g(x)$ the gluon PDFs. 

With these moment-space Wilson coefficients, it can be shown that the real parts of the gCFFs read,
\begin{align}
\begin{split}
    \text{Re}\mathcal{H}_{g}(\xi,t)= \sum_{n=0}^{\infty}\frac{1}{\xi^{2n+2}} \bar{C}\ui{2n+1}_g \bar{H}_g\ui{2n+1}(\xi,t)\ ,
\end{split}
\end{align}
as presented in the letter, noting that only odd moments contribute as gluon GPDs are symmetric, and similarly for the singlet quark GPDs as well. We also note that this expansion does not provide an explicit expression of the imaginary part of the amplitudes, which are generally non-zero for $\xi< 1$. The underlying reason is that this summation is divergent (as we discussed in the previous work~\cite{Guo:2023qgu}, it is an asymptotic series), so the imaginary part emerges only after the analytical continuation of this formal summation.

\begin{figure}[t]
    \includegraphics[width=0.98\textwidth]{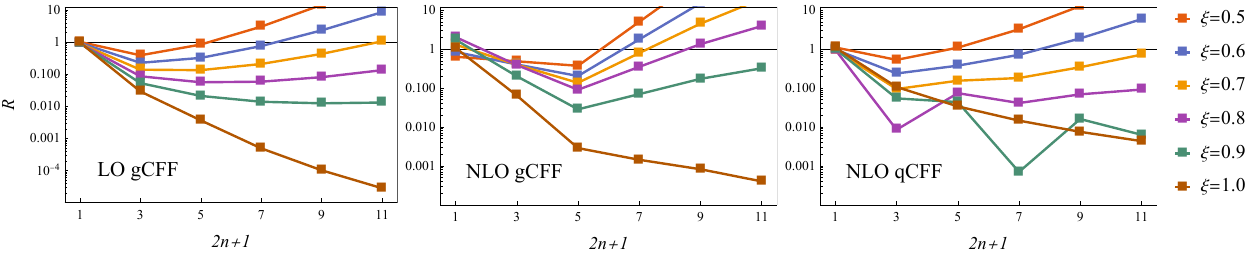}
    \caption
    {\raggedright The relative contributions of conformal moments to the real part of CFFs for LO gluon (left), NLO gluon (middle) and NLO quark (right). The asymptotic expansions generally hold, but for the less singular Wilson coefficients, it appears more stable. The NLO gluon Wilson coefficients are the most singular one, which suffer from numerical errors that might cause the fluctuation of the contribution of each moment. }
    \label{fig:asymplot}
\end{figure}

Then, such formal summations can be used to estimate the real part of the q/gCFFs in the large-$\xi$ limit according to the superasymptotic approximation. Previously, we studied the asymptotic behavior with LO Wilson coefficients~\cite{Guo:2023qgu}, and here we extend to the NLO.
In Fig. \ref{fig:asymplot}, we show the relative contributions of conformal moments to the real part of the LO and NLO q/gCFFs, with the ratio $R$ defined as,
\begin{equation}
    R\ui{2n+1}\equiv \frac{\xi^{-2n-2} \bar{C}\ui{2n+1}_g \bar{H}_g\ui{2n+1}(\xi,t)}{\text{Re}\mathcal{H}_{g\rm{C}}(\xi,t)} 
\end{equation}
where typical behaviors of asymptotic expansions are indeed seen for all three of them --- each term in the series starts by decreasing but eventually diverge. The leading moment with $2n+1=1$ contributes dominantly for large $\xi$ close to 1. In the example, we use the double distribution model of both quark and gluon GPDs~\cite{Radyushkin:1998bz} as in the previous work, where the quark and gluon PDFs are taken from the CT18 global analysis~\cite{Hou:2019efy}. Note that only valence up and down quark PDFs that dominate at large $x$ are used for simplicity, though the asymptotic behaviors should be generally independent of models, as found previously~\cite{Guo:2023qgu}.

In Fig. \ref{fig:asymplot}, it can be read that the contribution of the next-to-leading moment with $2n+1=3$ is around 30\% of the leading one for $\xi \gtrsim 0.5$ at LO, whereas at NLO, it rises to $\mathcal{O}(1)$ for the gluonic one according to the middle plot therein. However, since the NLO Wilson coefficient is additionally suppressed by $\alpha_S$ which is around $0.3$ at $\mu=m_c$. We estimate the theoretical uncertainties from leading-moment approximation to be $30$\% for $\xi\gtrsim0.5$. We also note that the contribution from imaginary part was found comparable to the truncation errors from the previous analysis~\cite{Guo:2023qgu}.  As all these extra contributions will be enhanced when $\xi$ get smaller, we will exclude the data for $\xi<0.5$ in this analysis to avoid the breakdown of large-$\xi$ expansion .

\section{Summary of Bayesian inference results}

In this section, we present the detailed results from Bayesian inference that are discussed in the letter. All parameters are presented at the scale $\mu=m_c$ with NLO GPD evolution, different from the conventional choice of $\mu=2$ GeV.

\subsection{Bayesian analysis results with the Lattice GFFs only}
\label{append:lattice}

\begin{figure}[th]
    \includegraphics[width=0.4\textwidth]{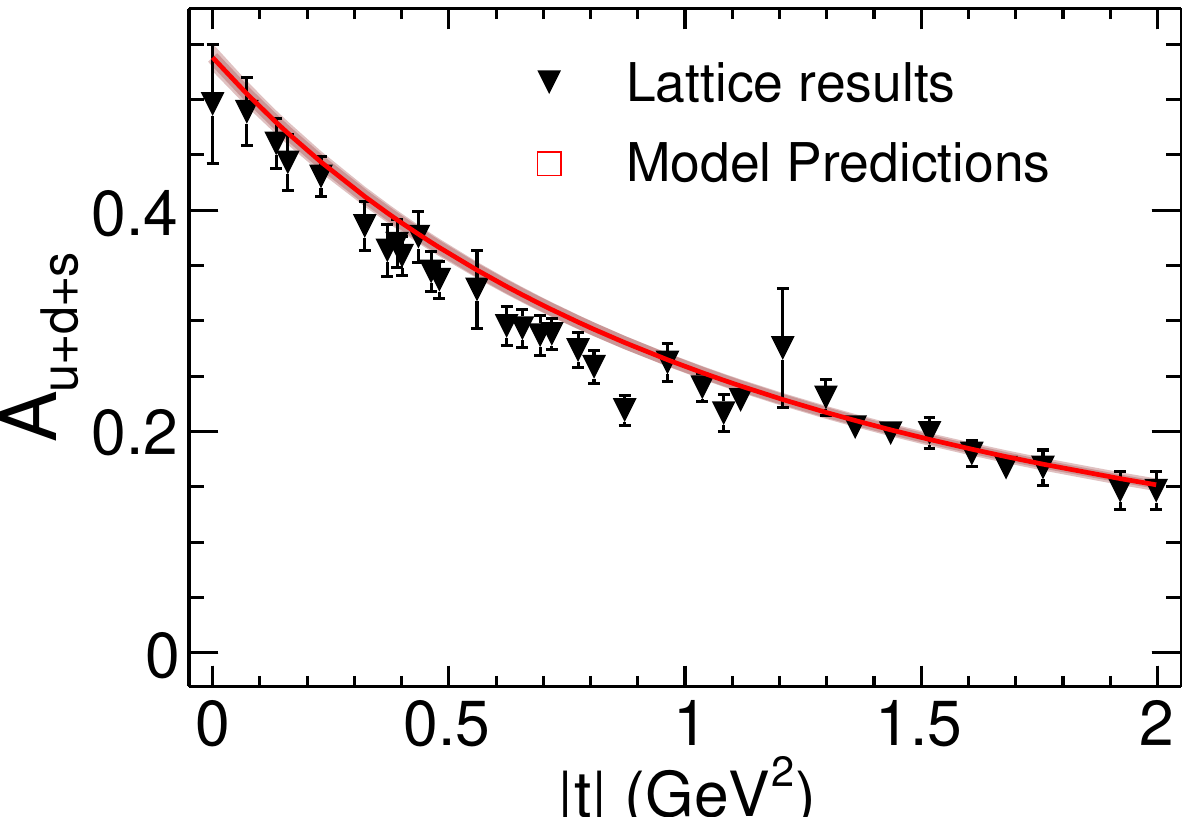}
    \includegraphics[width=0.4\textwidth]{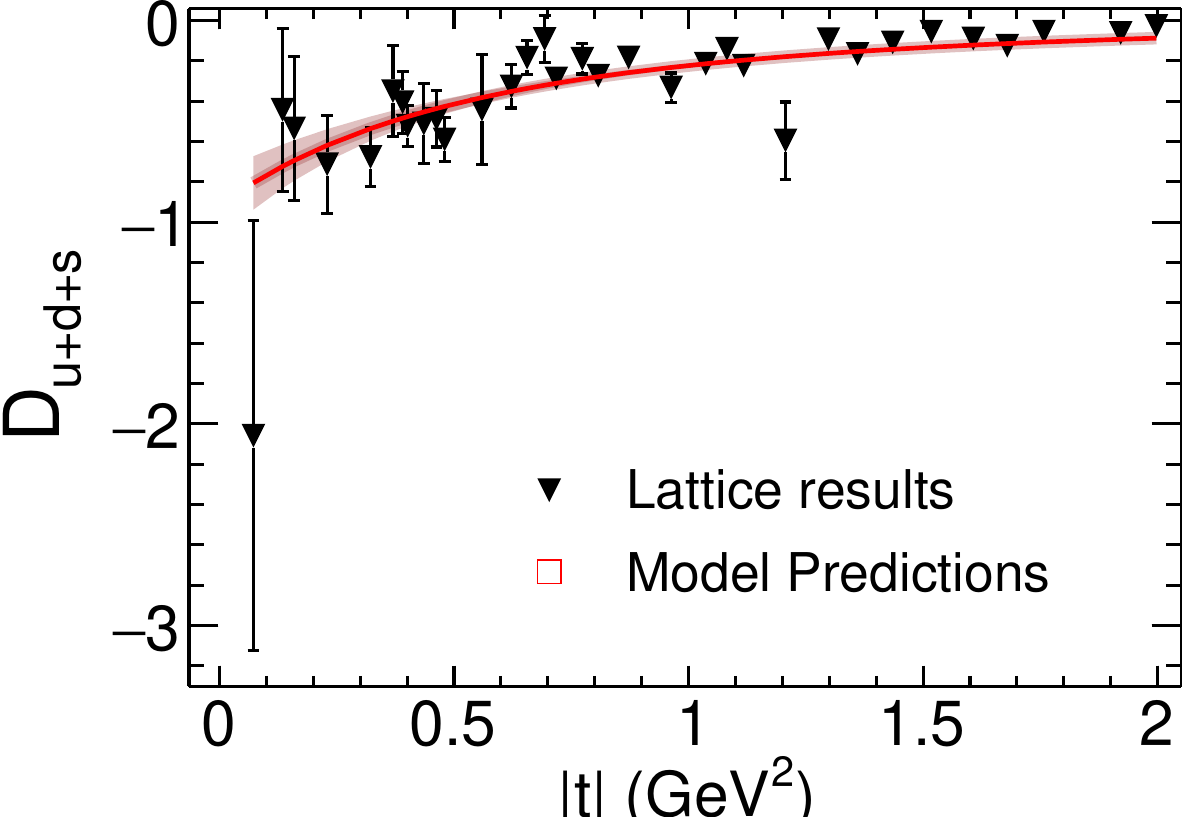}
    \includegraphics[width=0.4\textwidth]{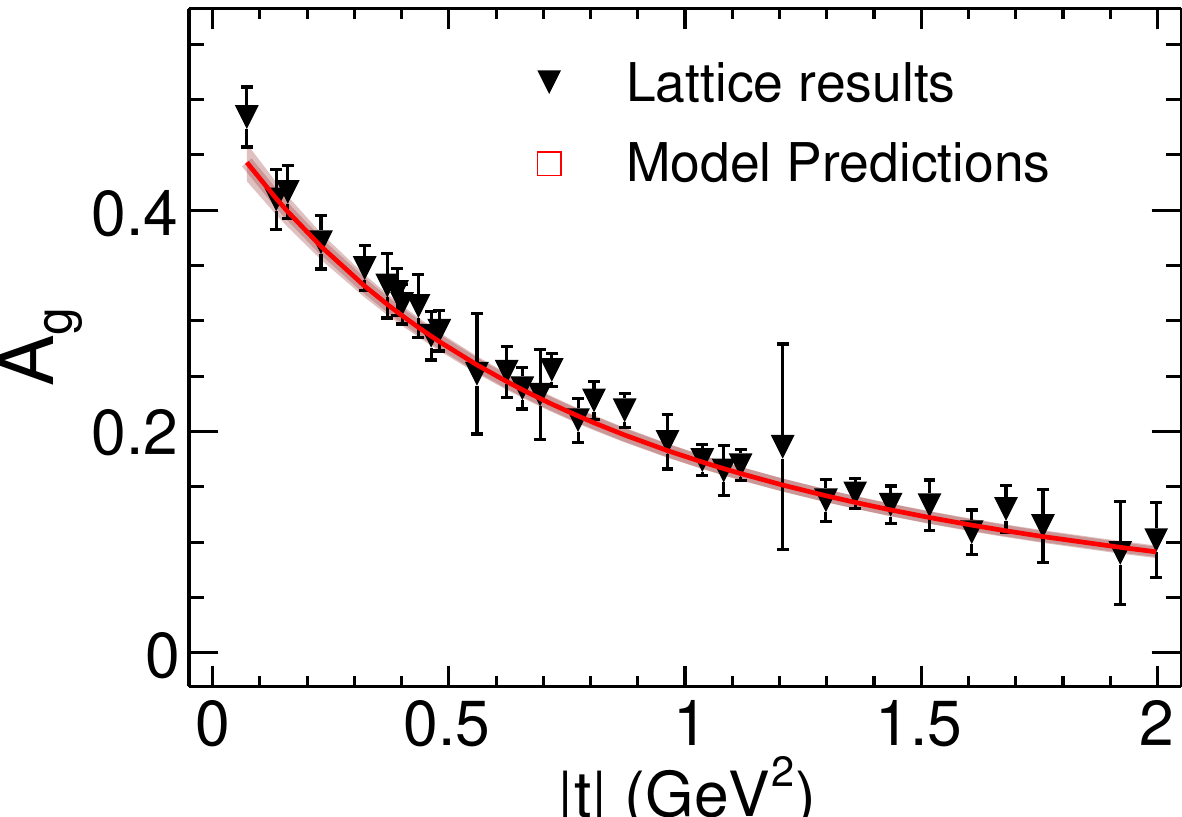}
    \includegraphics[width=0.4\textwidth]{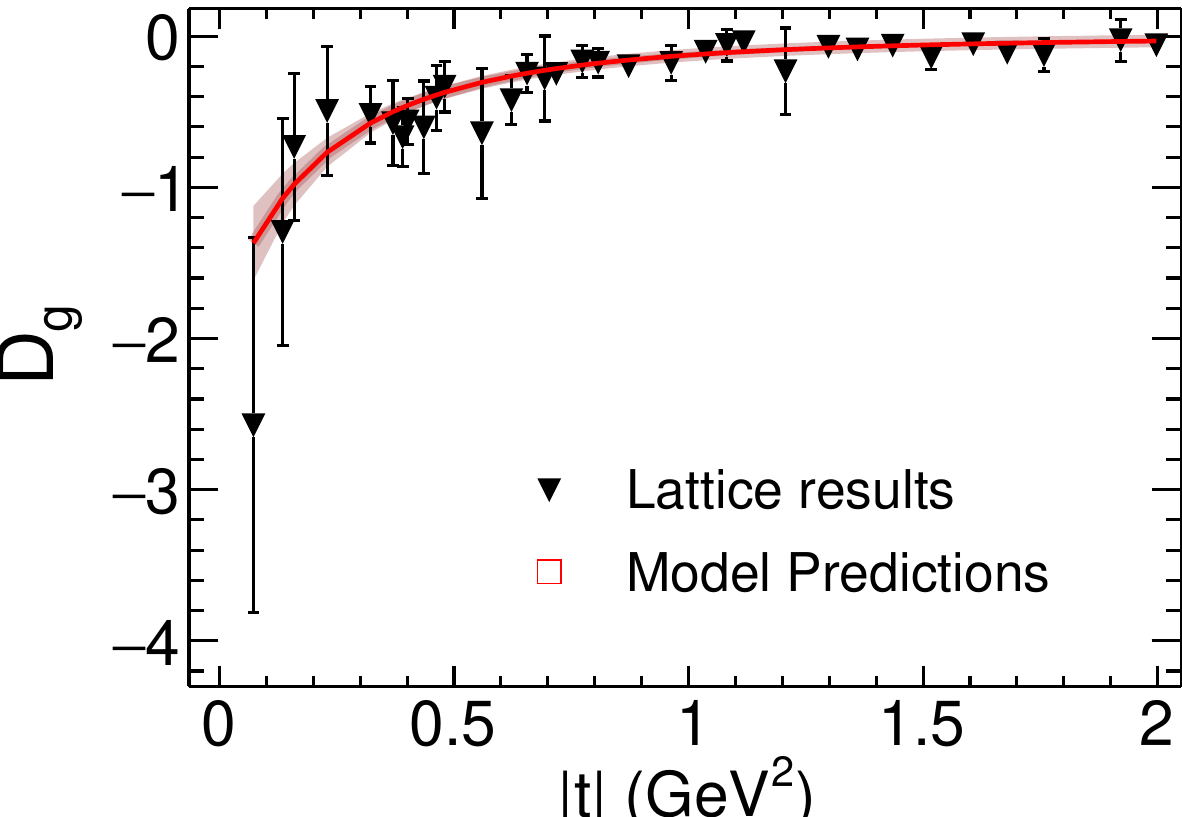}
    \caption
    { The model predictions with the posterior distributions of parameters fitting the Lattice results only.}
    \label{fig:bayes_lat_only_comp}
\end{figure}

\begin{figure}[th]
    \includegraphics[width=0.9\textwidth]{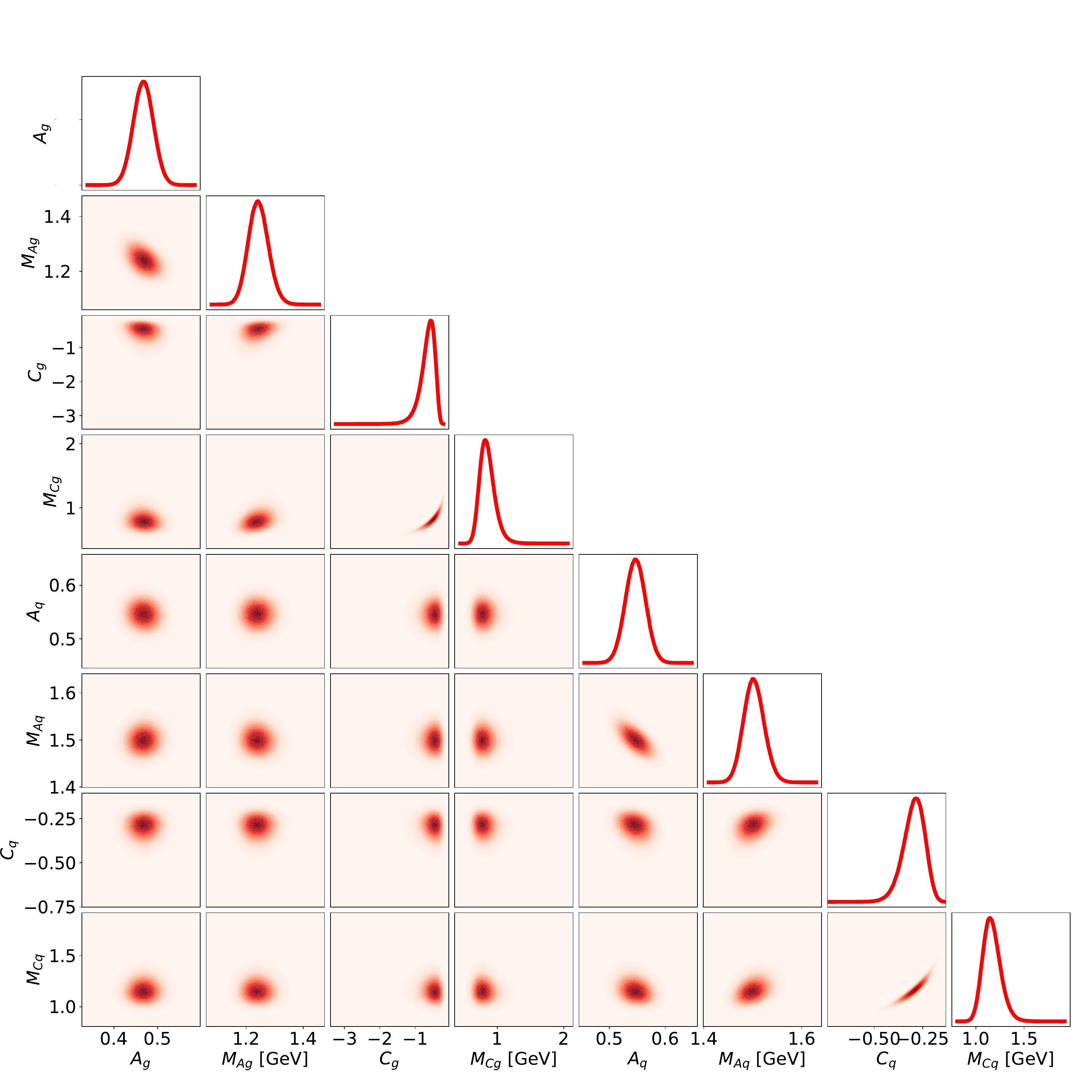}
    \caption
    { Bayesian posterior distributions of the model parameters by fitting the Lattice data only. The diagonal panels show the probability distributions for individual parameters, and off-diagonal panels illustrate their pairwise correlations.    }
    \label{fig:bayes_lat_only_post}
\end{figure}

First, we present the Bayesian analysis with only the lattice GFFs as the baseline of comparison~\cite{Hackett:2023rif}. Since the lattice GFFs are well constrained, the parameters are supposed to peak very well with the simple dipole/tripole fit. As shown in FIG. \ref{fig:bayes_lat_only_comp}, the lattice GFFs are indeed well described by the simple dipole/tripole parameterization, and the Bayesian posterior distributions are shown in FIG. \ref{fig:bayes_lat_only_post}. The posterior distributions all have sharp peaks, indicating that the parameters are well constrained by the lattice GFFs, unsurprisingly. Note the slightly different notation $D_{q/g}(t)\equiv 4 C_{q/g}(t)$ from the letter to be consistent with the presentation in~\cite{Hackett:2023rif}.

\subsection{Bayesian analysis results with the Lattice GFFs and experimental data with NLO formulae}

\begin{figure}[ht]
    \includegraphics[width=0.4\textwidth]{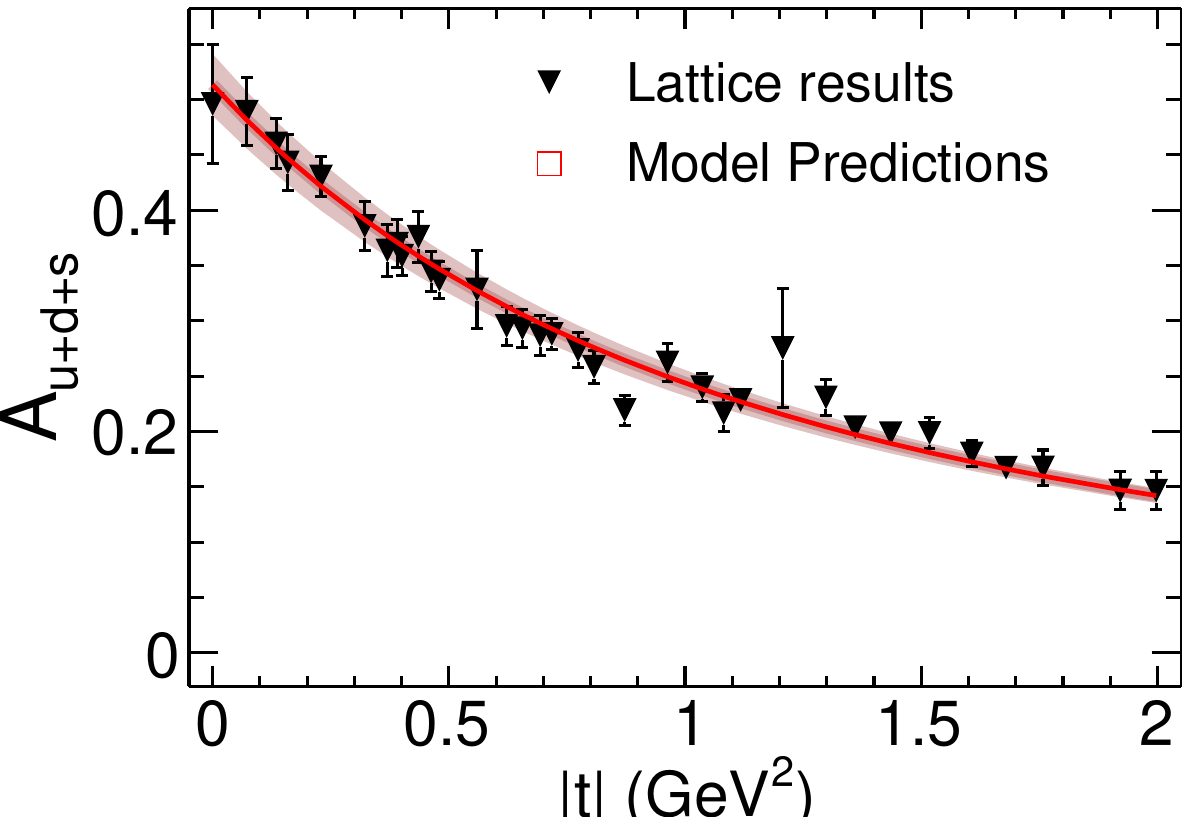}
    \includegraphics[width=0.4\textwidth]{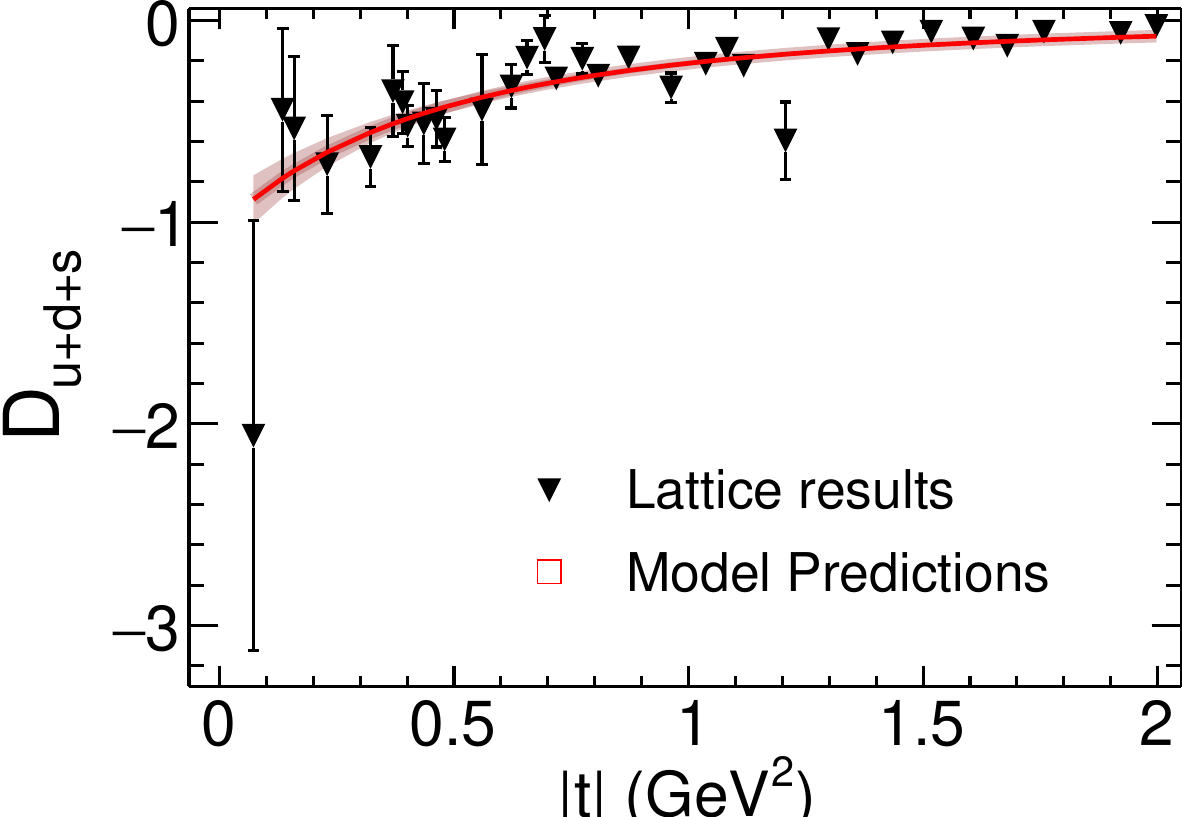}
    \includegraphics[width=0.4\textwidth]{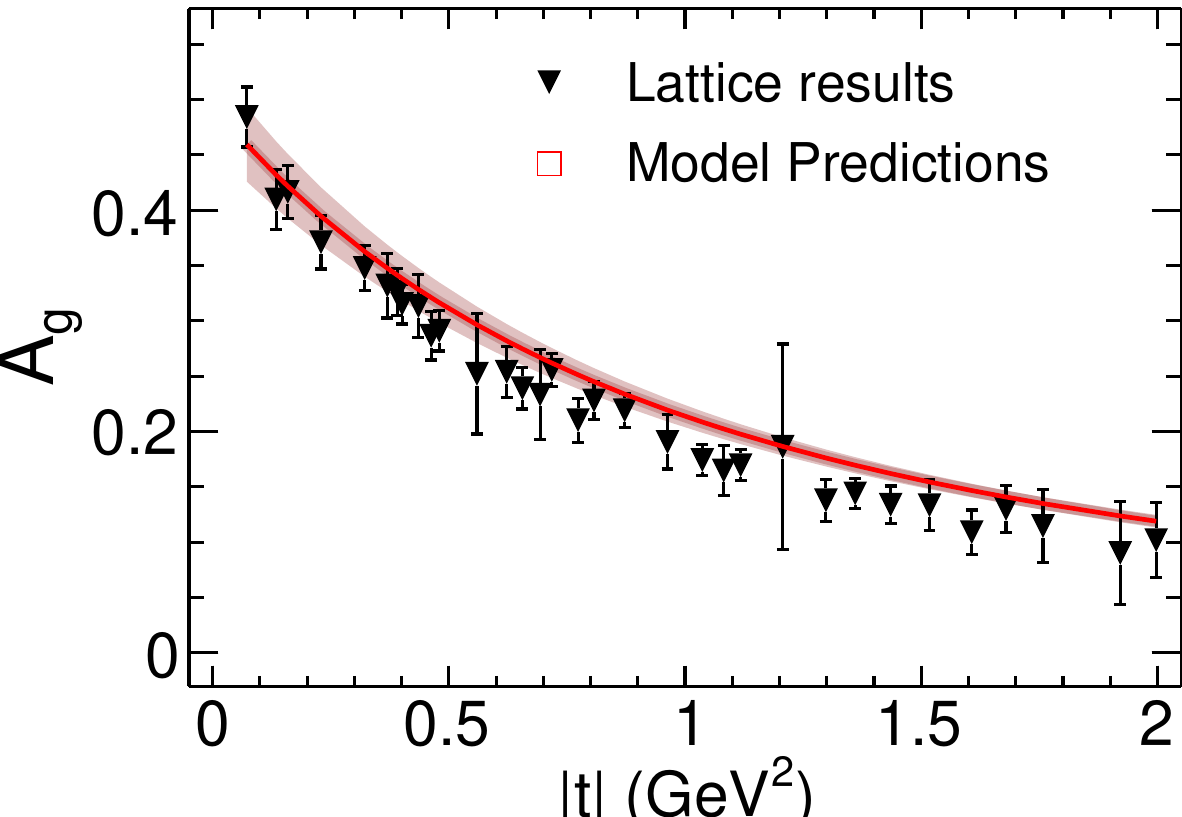}
    \includegraphics[width=0.4\textwidth]{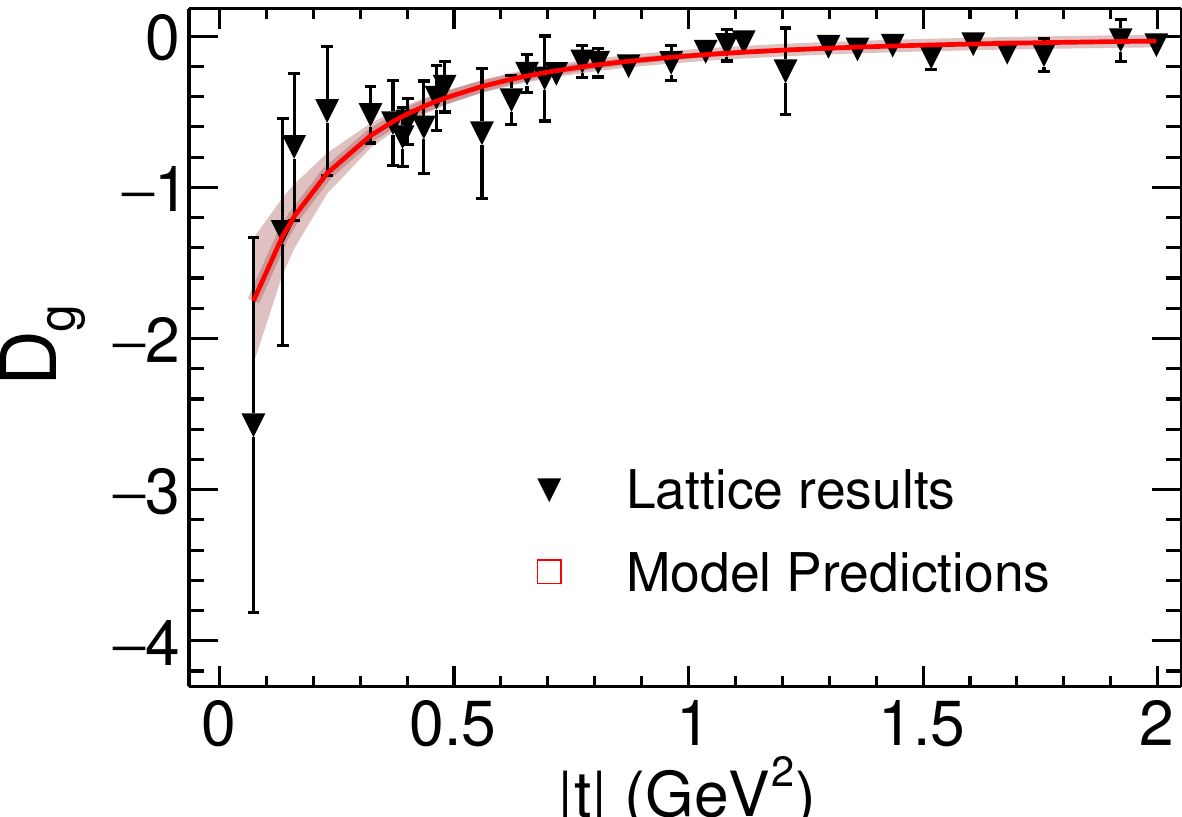}
    \includegraphics[width=0.4\textwidth]{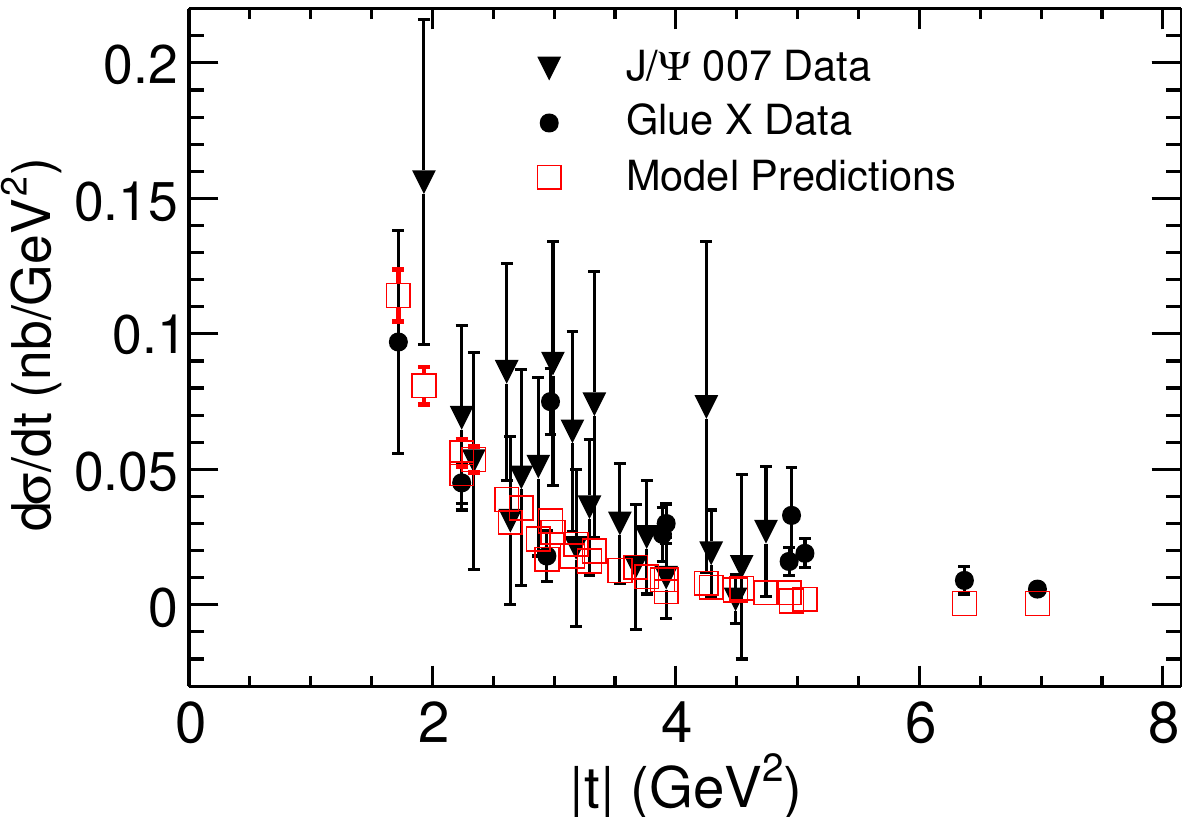}
    \caption
    { The model predictions with the posterior distributions of parameters fitting the Lattice results and experimental data simultaneously with NLO formulae.}
    \label{fig:latPlusexpnlo_comp}
\end{figure}

\begin{figure}[ht]
    \includegraphics[width=0.9\textwidth]{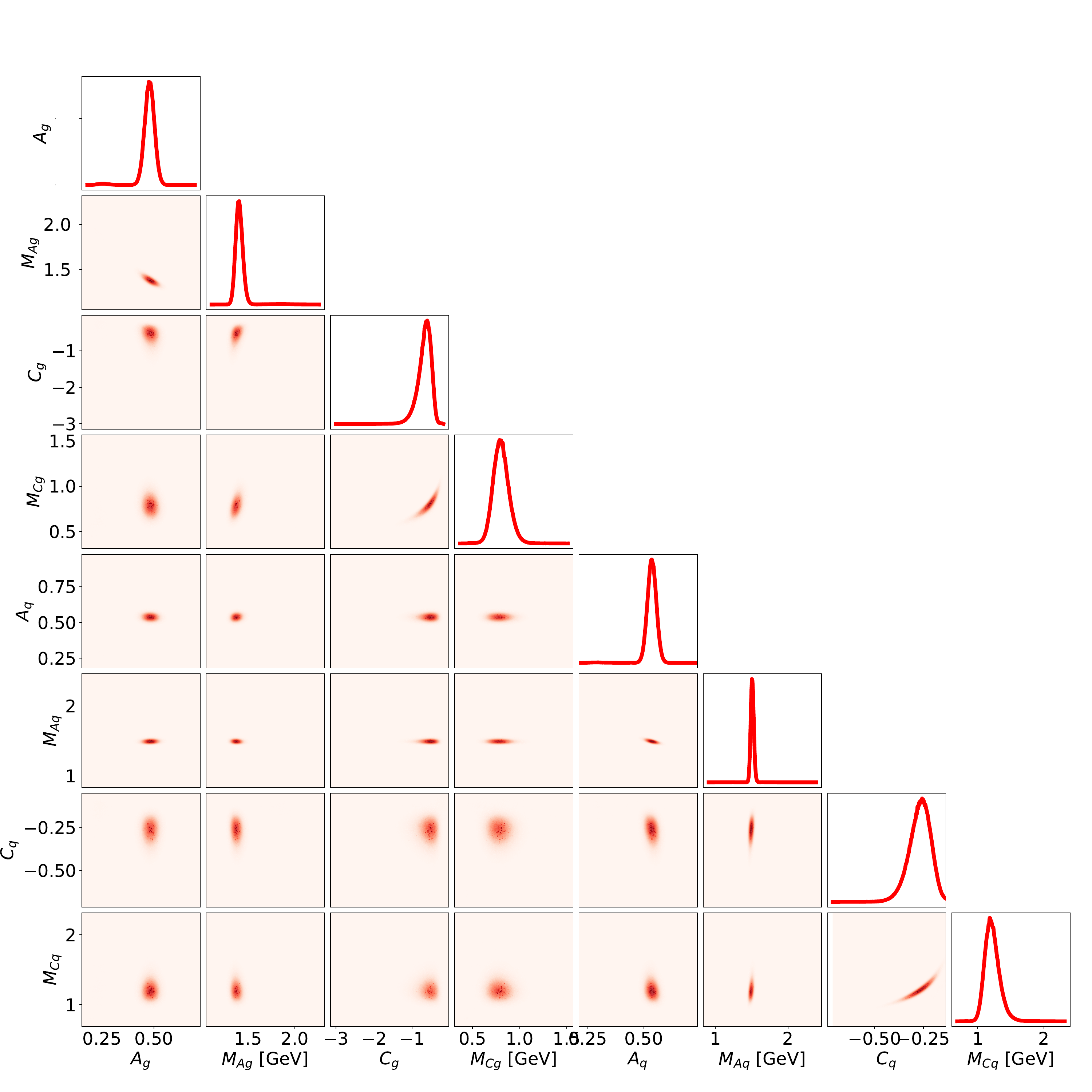}
    \caption
    { Bayesian posterior distributions of the model parameters by fitting the Lattice data and experimental data with NLO formulae simultaneously. The diagonal panels show the probability distributions for individual parameters, and off-diagonal panels illustrate their pairwise correlations.    }
    \label{fig:latPlusexpnlo_post}
\end{figure}

In FIG. \ref{fig:latPlusexpnlo_comp}, we also present the fit to combined lattice GFFs and  experimental data with NLO formulae, and the Bayesian posterior distributions are given in FIG. \ref{fig:latPlusexpnlo_post}. Due to the drastically larger statistics of the lattice GFFs compared to that of the experimental data, the posterior distributions do not differ much from the previous ones with lattice data only. But still, the experimental data provides rather non-trivial constraints on the gluonic $C_{g}(t)$, enhancing its sensitivity and leading to a smaller $|C_{g}(0)|$ than the lattice determination. This represents the great potential to constrain the gluonic GFFs via future high-precision data at large $\xi$. We also note that it is rather non-trivial that these posterior distributions can describe the data well so that the lattice GFFs are compatible with the experimental cross-sections, as shown in the bottom plot of FIG. \ref{fig:latPlusexpnlo_comp}. Particularly, this will not be the case when including the lower-$\xi$ data, where the theoretical uncertainties get out of control and this framework of large-$\xi$ expansion ceases to work, as we discussed above.

\subsection{Bayesian analysis results with only experimental data with NLO formulae}

\begin{figure}[ht]
    \includegraphics[width=0.45\textwidth]{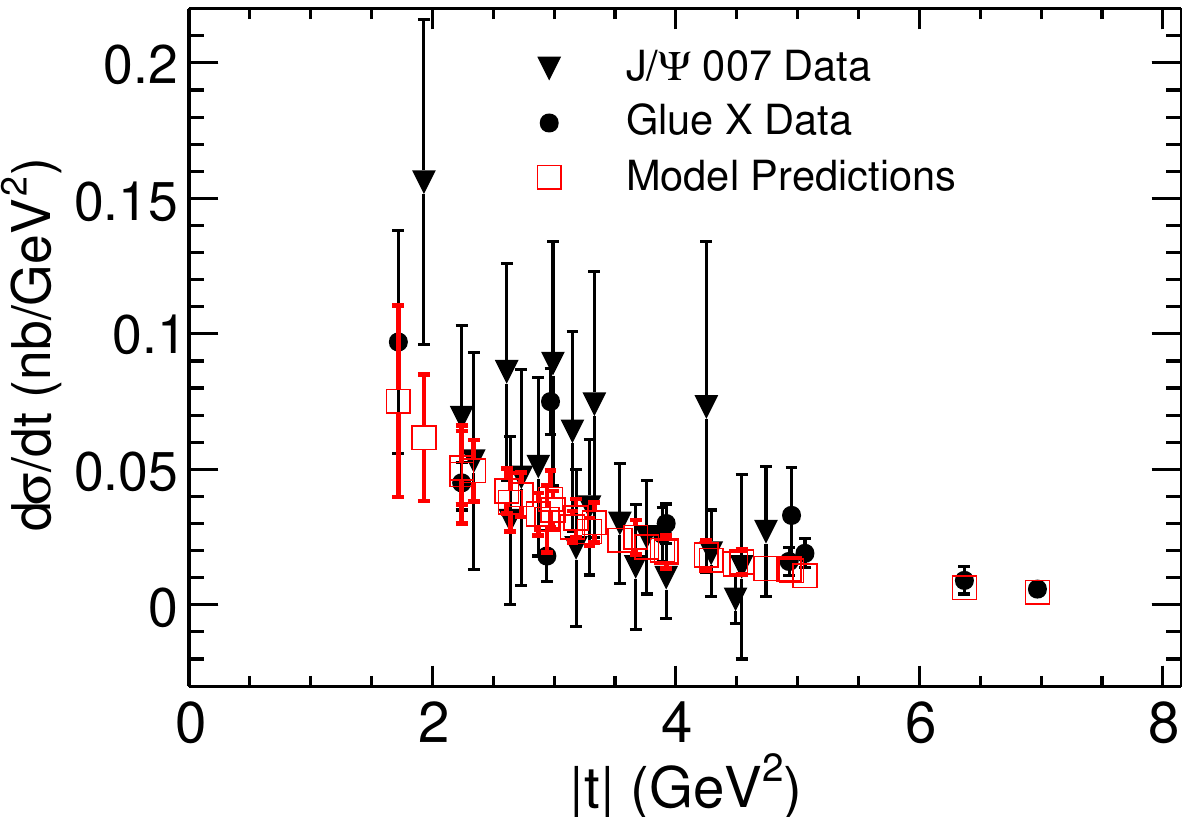}
    \caption
    { The model predictions with the posterior distributions of parameters fitting the experimental data only with NLO formulae, fixed $A_g = 0.4, A_q = 0.6$ and 40\% theoretical uncertainties in q/gCFFs.}
    \label{s.fig:expnlo_only_comp}
\end{figure}

\label{append:exponly}
\begin{figure}[ht]
    \includegraphics[width=0.8\textwidth]{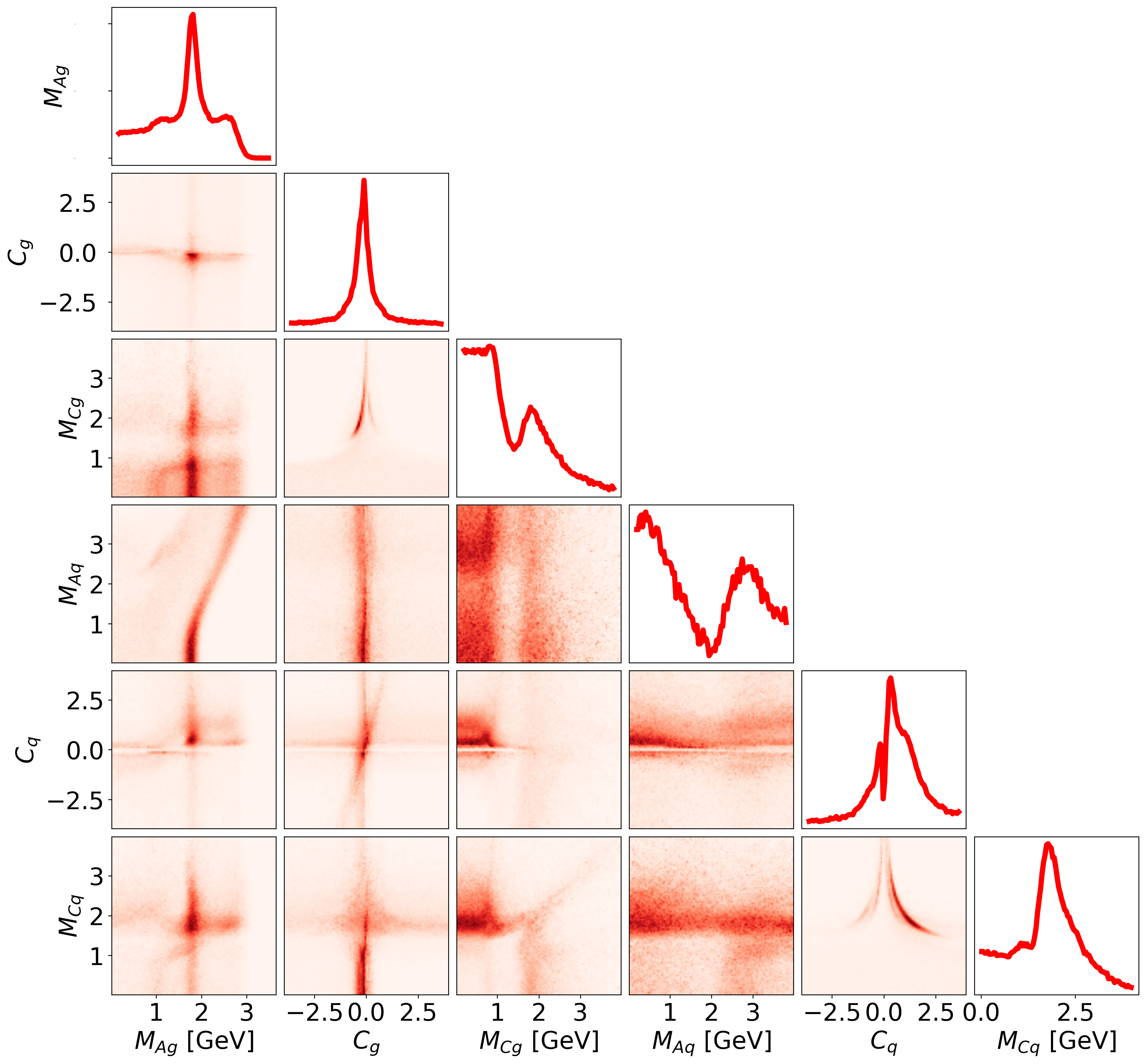}
    \caption
    { Bayesian posterior distributions of the model parameters by fitting the  experimental data only with NLO formulae. The diagonal panels show the probability distributions for individual parameters, and off-diagonal panels illustrate their pairwise correlations with fixed $A_g = 0.4, A_q = 0.6$ and 40\% theoretical uncertainties in q/gCFFs.    }
    \label{fig:expnlo_only_post}
\end{figure}

Finally, we also show the detailed Bayesian results with experimental data only. The experimental data are obviously well described with the parameterization of the form factors, as shown in FIG. \ref{s.fig:expnlo_only_comp}. On the other hand, the lack of statistics and precision of these large-$\xi$ data indicates that the parameters will not be well determined from just the experimental data, which is shown to be the case according to Bayesian posterior distributions of the parameters shown in FIG. \ref{fig:expnlo_only_post}. Nevertheless, many posterior distributions of the parameters appear to peak, indicating non-trivial knowledge of the GFFs, particularly the $C_{q}(0)$ and $C_{g}(0)$ form factors, can be learned from the experimental data. The major findings have been discussed in the letter.

\end{document}